%% file: paper.tex
\pdfoutput=1
\documentclass[
natbib,10pt]{TOOLS/sigplanconf}
\input TOOLS/hheader.tex

\begin{document}

\title{
  Ranking Catamorphisms and Unranking Anamorphisms on Hereditarily Finite
  Datatypes
}
\subtitle{-- unpublished draft --}
           
\authorinfo{Paul Tarau}
   {Department of Computer Science and Engineering\\
   University of North Texas}
   {\em tarau@cs.unt.edu}

\maketitle

\date{}

\begin{abstract}
Using specializations of {\em unfold} and {\em fold} on a 
generic tree data type we derive {\em unranking} and {\em ranking} 
functions providing natural number encodings 
for various Hereditarily Finite datatypes. 

In this context, we interpret unranking operations as instances of a generic
{\em anamorphism} and ranking operations as instances of the corresponding
{\em catamorphism}.

Starting with Ackerman's Encoding from Hereditarily Finite Sets to
Natural Numbers
we define {\em pairings} and {\em finite tuple encodings} that provide building
blocks for a theory of {\em Hereditarily Finite Functions}. 

The more difficult problem
of {\em ranking} and {\em unranking} {\em Hereditarily Finite Permutations} is
then tackled using Lehmer codes and factoradics.

The self-contained source code of the paper, as generated from a
literate Haskell program, is available at
\url{http://logic.csci.unt.edu/tarau/research/2008/fFUN.zip}.

\end{abstract}

\section{Introduction}
This paper is an exploration with functional programming tools of {\em ranking}
and {\em unranking} problems on finite functions and bijections and their related
hereditarily finite universes. 
The {\em ranking problem} for a family of
combinatorial objects is finding a unique natural number associated to it,
called its {\em rank}.
The inverse {\em unranking problem} consists of generating a unique
combinatorial object associated to each natural number. 

The paper is organized as follows: section \ref{unrank} 
introduces generic ranking/unranking framework
parameterized by bijective transformers and terminating
conditions based on {\em urelements}, section \ref{ack} introduces Ackermann's
encoding and its inverse as instances of the framework. 
After discussing some classic pairing functions, section \ref{pairings} introduces new 
pairing/unpairing and tuple operations on natural numbers and uses them for encodings 
of finite functions (section \ref{fun}), resulting in encodings for 
``Hereditarily Finite Functions" (section \ref{hff}).
Ranking/unranking of permutations and Hereditarily Finite
Permutations as well as Lehmer codes and factoradics are covered in 
section \ref{perm}.
Sections \ref{related} and \ref{concl} discuss related work, 
future work and conclusions.

The paper is part of a larger effort to cover in a declarative programming 
paradigm, arguably more elegantly, some fundamental combinatorial generation 
algorithms along the lines of \cite{knuth06draft}. The practical 
expressiveness of functional programming languages (in particular Haskell) 
are put at test in the process. 

While the main focus of the paper was testdriving Haskell on
the curvy tracks of non-trivial combinatorial generation problems, 
we have bumped, somewhat accidentally, into
making a few new contributions to the field as such, 
that could be easily blamed on the quality of the
vehicle we were testdriving:
\begin{enumerate}
  \item the three ranking/unranking algorithms from finite functions to natural
  numbers are new
  \item the universe of Hereditarily Finite Functions, as a functional
  analogue of the well known universe of Hereditarily Finite Sets is new
\item 
  the universe of Hereditarily Finite Permutations, as an
  analogue of the well known universe of Hereditarily Finite Sets is new  
\item 
  the natural number tupling/untupling functions are
  new
\item the ranking/unranking algorithm for permutations of 
      arbitrary sizes is new (although it is based on a known Lehmer code-based
      algorithm for permutations of fixed size)
\item the catamorphism/anamorphism view of ranking/unranking functions is new
and it is likely to be reusable for various families of combinatorial generation
problems
\end{enumerate}

Through the paper, we will use the following set of primitive arithmetic functions:
\begin{code}
double n = 2*n
half n = n `div` 2
exp2 n = 2^n
\end{code}
together with {\tt succ}, {\tt pred}, {\tt even}, {\tt odd} and {\tt sum}
Haskell functions, to emphasize that this subset is easily hardware
implementable (by only using boolean operations, shifts and adders) 
and that these functions also have {\tt
O(log n)} or better software implementations for integers of (arbitrary) length
{\tt n}.

When possible, we will use point-free notations (unnecessary function arguments
omitted) to emphasize the generic function composition dataflow. 

As we have put
significant effort to ensure that all our types can be inferred, we will omit
type declarations, with apologies to the type-curious reader, who can have
them displayed as needed, while interacting with the Haskell
sources of the paper available online.

\section{Generic Unranking and Ranking with Higher Order Functions} \label{unrank}

We will use, through the paper, a generic ``rose tree'' type {\tt T} distinguishing 
between atoms tagged with {\tt A}) and subforests (tagged with {\tt F}).
\begin{code}
data T a = A a | F [T a] deriving (Eq,Ord,Read,Show)
\end{code}
Atoms will be mapped to natural numbers in {\tt [0..ulimit-1]}. When {\tt
ulimit} is fixed, we denote this set $A$. We denote $Nat$ the set of natural numbers and $T$ the set of trees 
of type $T$ with atoms in $A$.

The unranking operation is seen here as an instance of a generic
anamorphism mechanism {\em unfold}, while the ranking operation is seen as an
instance of the corresponding catamorphism {\em fold}
\cite{DBLP:journals/jfp/Hutton99,DBLP:conf/fpca/MeijerH95}.

\paragraph{Unranking} As an adaptation of the {\em unfold} 
operation, natural numbers will be mapped to elements of $T$ with a generic 
higher order function {\tt unrank\_ ulimit f}, defined from $Nat$ to $T$, 
parameterized by the the natural number {\tt ulimit} 
and the transformer function {\tt f}:

\begin{code}
unrank_ ulimit _ n | (n<ulimit) && (n>=0) = A n
unrank_ ulimit f n | n >=ulimit = 
  (F (map (unrank_ ulimit f) (f  (n-ulimit))))
\end{code}

A global constant {\tt default\_ulimit} will be used 
through the paper to fix the default range of 
atoms, allowing us to work with a default {\tt unrank} function: 
\begin{code}
default_ulimit = 0

unrank = unrank_ default_ulimit
\end{code}

\paragraph{Ranking} Similarly, as an adaptation of {\em fold}, generic inverse 
mappings {\tt rank\_ ulimit} and {\tt rank}) from $T$ to $Nat$ are defined as:
\begin{code}
rank_ ulimit _ (A n) | (n<ulimit) && (n>=0) = n
rank_ ulimit g (F ts) = 
  ulimit+(g (map (rank_ ulimit g) ts))

rank = rank_ default_ulimit
\end{code}
Note that the guard in the second definition simply states correctness constraints ensuring 
that atoms belong to the same set $A$ for {\tt rank\_} and {\tt unrank\_}. This ensures 
that the following holds:

\begin{prop}
If the transformer function $f:Nat \rightarrow [Nat]$ is a bijection with inverse g, 
such that $n \geq ulimit \wedge f(n)=[n_0,...n_i,...n_k] \Rightarrow n_i<n$, then {\tt unrank} 
is a bijection from $Nat$ to $T$, with inverse {\tt rank} and the recursive computations 
of both functions terminate in a finite number of steps.

{\em Proof:} by induction on the structure of $Nat$ and $T$, using the fact that {\tt map} 
preserves bijections.
\end{prop}

{\em Ranking} functions can be traced back to G\"{o}del numberings
\cite{Goedel:31,conf/icalp/HartmanisB74} associated to formulae. 
Together with their inverse {\em unranking} functions they are also 
used in combinatorial generation
algorithms \cite{conf/mfcs/MartinezM03,knuth06draft}.

\section{Hereditarily Finite Sets and Ackermann's Encoding} \label{ack}

While the Universe of Hereditarily Finite Sets is best known as a model of the 
Zermelo-Fraenkel Set theory with the Axiom of Infinity  replaced by its 
negation \cite{finitemath,DBLP:journals/jct/MeirMM83}, it has been the object 
of renewed practical interest in various fields, from representing structured data 
in databases \cite{DBLP:conf/foiks/LeontjevS00} to reasoning with sets and set constraints 
 \cite{dovier00comparing,DBLP:journals/tplp/PiazzaP04}.

\subsection{Ackermann's Encoding} 

The Universe of Hereditarily Finite Sets is built from the empty set (or a set of {\em Urelements}) 
by successively applying powerset and set union operations.

A surprising bijection, discovered by Wilhelm Ackermann in 1937 \cite{ackencoding,abian78,kaye07} 
maps Hereditarily Finite Sets ($HFS$) to Natural Numbers ($Nat$):

\vskip 0.5cm
$f(x)$ = {\tt if} $x=\{\}$ {\tt then} $0$ {\tt else} $\sum_{a \in x}2^{f(a)}$
\vskip 0.5cm

Assuming $HFS$ extended with {\em Urelements} (objects not containing any
elements) our generic ``rose tree'' representation can be used for Hereditarily
Finite Sets, with {\em Urelements} seen as atoms, i.e. Natural Numbers in {\tt
[0..ulimit-1]}. The constructor {\tt A a} marks {\em Urelements} of type {\tt a} (usually the arbitrary 
length Integer type in Haskell) and the constructor {\tt F} marks a list of 
recursively built $HFS$ type elements. Note that if no elements are used with the {\tt A} 
constructor, we obtain the ``pure'' $HFS$ universe with everything built out 
from the empty set represented as {\tt F []}.

Let us note that Ackermann's encoding can be seen as the recursive application of 
a bijection {\tt set2nat} from finite subsets of $Nat$ to $Nat$, that associates to 
a set of (distinct!) natural numbers a (unique!) natural number. 
With this representation, Ackermann's encoding from $HFS$ to $Nat$ {\tt hfs2nat} 
can be expressed in terms of our generic {\tt rank} function as:
\begin{code} 
hfs2nat = rank set2nat

set2nat ns = sum (map exp2 ns)
\end{code}

To obtain the inverse of the Ackerman encoding, let's first define the 
inverse {\tt nat2set} of the bijection {\tt set2nat}. It decomposes a natural number 
into a list of exponents of 2 (seen as bit positions equaling 1 in its bitstring 
representation, in increasing order).
\begin{code}
nat2set n = nat2exps n 0 where
  nat2exps 0 _ = []
  nat2exps n x = 
    if (even n) then xs else (x:xs) where
      xs=nat2exps (half n) (succ x)
\end{code}
The inverse of the  Ackermann encoding, with urelements in {\tt
[0..default\_ulimit-1]} and the empty set mapped to {\tt F []} is defined as
follows:
\begin{code}
nat2hfs = unrank nat2set
\end{code}
This definition is motivated by the fact that {\tt nat2hfs} and {\tt hfs2nat}
are obtained through recursive compositions of {\tt nat2set} and {\tt set2nat}, respectively. 
Generalizing the encoding mechanism to use
other  bijections with similar properties, 
naturally leads to the {\em anamorphism/catamorphism}
view of {\tt unrank/rank}.

The following proposition summarizes the results in this subsection:
\begin{prop}
Given id = $\lambda x.x$, the following function equivalences hold:

\begin{equation}
nat2set \circ set2nat \equiv id
\end{equation}

\begin{equation}
set2nat \circ nat2set \equiv id
\end{equation}

\begin{equation}
nat2hfs \circ hfs2nat \equiv id
\end{equation}

\begin{equation}
hfs2nat \circ nat2hfs \equiv id
\end{equation}
\end{prop}

\subsection{Combinatorial Generation as Iteration}
Using the inverse of Ackermann's encoding, the infinite stream $HFS$ 
can be generated simply by iterating over the infinite stream {\tt [0..]}:
\begin{code}
iterative_hfs_generator=map nat2hfs [0..]
\end{code}
\begin{verbatim}
take 5 iterative_hfs_generator
 [F [],F [F []],F [F [F []]],
    F [F [],F [F []]],F [F [F [F []]]]]
\end{verbatim}

One can try out {\tt nat2hfs} and its inverse {\tt hfs2nat} and 
print out a canonical string representation of $HFS$ with the 
{\tt setShow} functions given in Appendix:
\begin{verbatim}
nat2hfs 42
  F [F [F []],F [F [],F [F []]],
     F [F [],F [F [F []]]]]
hfs2nat (nat2hfs 42)
  42
setShow 42
  "{{{}},{{},{{}}},{{},{{{}}}}}"
\end{verbatim}
\noindent Note that {\tt setShow n} will build a string 
representation of $n \in Nat$, implicitly ``deforested'' as a $HFS$ with 
Urelements in {\tt [0..default\_ulimit-1]}. Figure \ref{f42} shows the directed
acyclic graph obtained by merging shared nodes in the {\em rose tree} 
representation of the $HFS$ 
associated to a natural number (with arrows pointing from sets to their elements).
\FIG{f42}{Hereditarily Finite Set associated to 42}{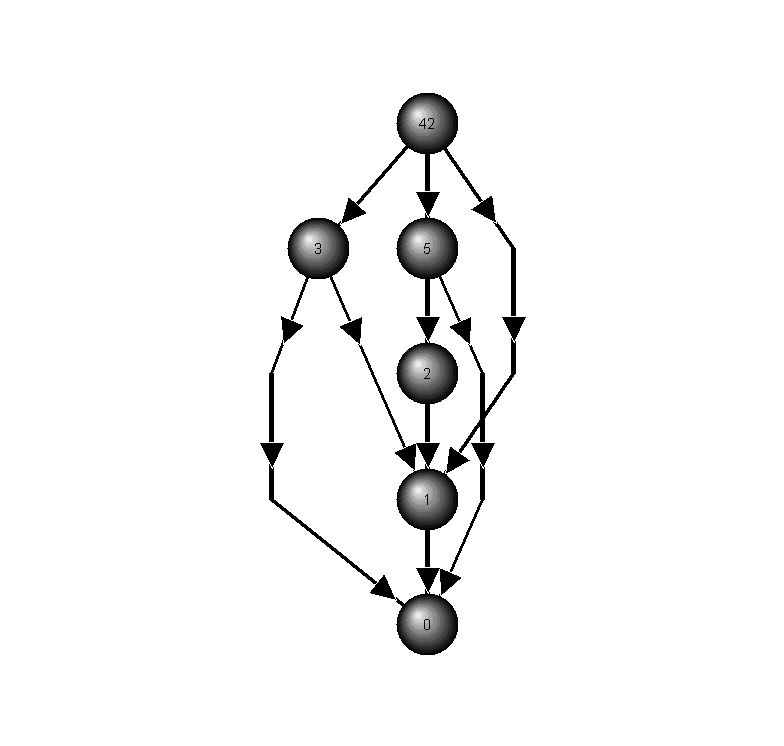}

\section{Pairing Functions and Tuple Encodings} \label{pairings}

{\em Pairings} are bijective functions $Nat \times Nat \rightarrow Nat$.  
Following the classic notation for pairings of \cite{robinson50}, 
given the pairing function $J$, its left and right inverses $K$ and $L$ are such that

\begin{equation}
J(K(z),L(z))=z
\end{equation}

\begin{equation}
K(J(x,y))=x
\end{equation}

\begin{equation} 
L(J(x,y))=y 
\end{equation}

We refer to  \cite{DBLP:journals/tcs/CegielskiR01} for a typical use 
in the foundations of mathematics and to \cite{DBLP:conf/ipps/Rosenberg02a} 
for an extensive study of various pairing functions and their computational properties.
We will start by overviewing two classic pairing functions.

\subsection{Cantor's Pairing Function}
Cantor's geometrically inspired pairing function (also present in earlier work
by Cauchy) is defined as:
\begin{code}
nat_cpair x y = (x+y)*(x+y+1) `div` 2+y
\end{code}
As the following example shows, it grows symmetrically in both arguments:
\begin{verbatim}
[nat_cpair i j|i<-[0..3],j<-[0..3]]
  [0,2,5,9,1,4,8,13,3,7,12,18,6,11,17,24]
\end{verbatim}

\subsection{The Pepis-Kalmar-Robinson Pairing Function}
An interesting pairing function {\em  asymmetrically growing, faster on the first argument}, 
is the function {\bf pepis\_J} and its left and right unpairing companions 
{\bf pepis\_K} and {\bf pepis\_L} that have been used, by Pepis, Kalmar and Robinson 
together with Cantor's functions, in some fundamental work on recursion theory, 
decidability and Hilbert's Tenth Problem in
\cite{pepis,kalmar1,kalmar2,kalmar3,robinson50,robinson55,robinson68a,robinsons68b,robinson67}.
The function {\bf pepis\_J}
combines two numbers reversibly by multiplying
a power of 2 derived from the first and
an odd number derived from the second:
\begin{equation}
f(x,y)=2^x*(2*y+1)-1
\end{equation}
Its Haskell implementation, together with its inverse is:
\begin{code}
pepis_J x y  = pred ((exp2 x)*(succ (double y)))

pepis_K n = two_s (succ n)

pepis_L n = half (pred (no_two_s (succ n)))
 
two_s n | even n = succ (two_s (half n))
two_s _ = 0

no_two_s n = n `div` (exp2 (two_s n))
\end{code}
This pairing function (slower in the second argument) works as follows:
\begin{verbatim}
pepis_J 1 10
  41
pepis_J 10 1
  3071
[pepis_J i j|i<-[0..3],j<-[0..3]]
  [0,2,4,6,1,5,9,13,3,11,19,27,7,23,39,55]
\end{verbatim}
As Haskell provides a built-in ordered pair, it is convenient to regroup {\tt J, K, L} 
as mappings to/from built-in ordered pairs:
\begin{code}
haskell2pepis (x,y) = pepis_J x y
pepis2haskell n = (pepis_K n,pepis_L n)
\end{code}

\subsection{The BitMerge Pairing Function} \label{BitMerge}

We will introduce here an unusually simple pairing function (that we
have found out recently as being the same as the one in defined 
in Steven Pigeon's PhD thesis on Data Compression \cite{pigeon}, page 114).
It provides compact representations for various constructs involving ordered pairs.

The bijection {\tt bitmerge\_pair} from $Nat \times Nat$ to $Nat$ 
and its inverse {\tt bitmerge\_unpair} 
are defined as follows:
\begin{code}
bitmerge_pair (i,j) = 
  set2nat ((evens i) ++ (odds j)) where
    evens x = map double (nat2set x)
    odds y = map succ (evens y)
  
bitmerge_unpair n = (f xs,f ys) where 
  (xs,ys) = partition even (nat2set n)
  f = set2nat . (map half)
\end{code}
The function {\tt bitmerge\_pair} works by splitting a number's big endian bitstring 
representation into odd and even bits while its inverse {\tt bitmerge\_unpair}
blends the odd and even bits back together. With help of the function {\tt to\_rbits}
given in Appendix, that decomposes $n \in Nat$ into a list of bits (smaller
units first) one can follow what happens, step by step:
\begin{verbatim}
to_rbits 2008
  [0,0,0,1, 1,0,1,1, 1,1,1]
bitmerge_unpair 2008
  (60,26)
to_rbits 60
  [0,0, 1,1, 1,1]
to_rbits 26
  [0,1, 0,1, 1]  
bitmerge_pair (60,26)
  2008
\end{verbatim}

\begin{prop}
The following function equivalences hold:

\begin{equation}
bitmerge\_pair \circ bitmerge\_unpair \equiv id
\end{equation}

\begin{equation}
bitmerge\_unpair \circ bitmerge\_pair \equiv id
\end{equation}
\end{prop}

\subsection{Tuple Encodings as Generalized BitMerge} \label{tuple}
We will now generalize this pairing function to $k$-tuples and then we will 
derive an encoding for finite functions. 

The function {\tt to\_tuple:} $Nat \rightarrow Nat^k$ converts a natural 
number to a $k$-tuple by splitting its bit representation into $k$ groups, 
from which the $k$ members in the tuple are finally rebuilt. This operation 
can be seen as a transposition of a bit matrix obtained by expanding 
the number in base $2^k$:
\begin{code}  
to_tuple k n = map from_rbits (
    transpose (
      map (to_maxbits k) (
        to_base (exp2 k) n
      )
    )
  )
\end{code}
To convert a $k$-tuple back to a natural number we will merge their 
bits, $k$ at a time. This operation uses the transposition of a bit 
matrix obtained from the tuple, seen as a number in base $2^k$, 
with help from bit crunching functions given in Appendix:
\begin{code}
from_tuple ns = from_base (exp2 k) (
    map from_rbits (
      transpose (
        map (to_maxbits l) ns
      )
    )
  ) where 
      k=genericLength ns
      l=max_bitcount ns
\end{code}
The following example shows the decoding of {\tt 42}, its decomposition 
in bits (right to left), the formation of a $3$-tuple and the encoding 
of the tuple back to {\tt 42}.
\begin{verbatim}
to_rbits 42
  [0,1,0, 1,0,1]
to_tuple 3 42
  [2,1,2]
to_rbits 2
  [0,1]
to_rbits 1
  [1]
from_tuple [2,1,2]
  42
\end{verbatim}
Fig. \ref{3tuple} shows multiple steps of the same decomposition, 
with shared nodes collected in a DAG. Note that cylinders represent 
markers on edges indicating argument positions, the cubes indicate leaf 
vertices (0,1) and the small pyramid indicates the root where the expansion has started.
\VFIGS{3tuple}{Repeated 3-tuple expansions}{42}{2008}{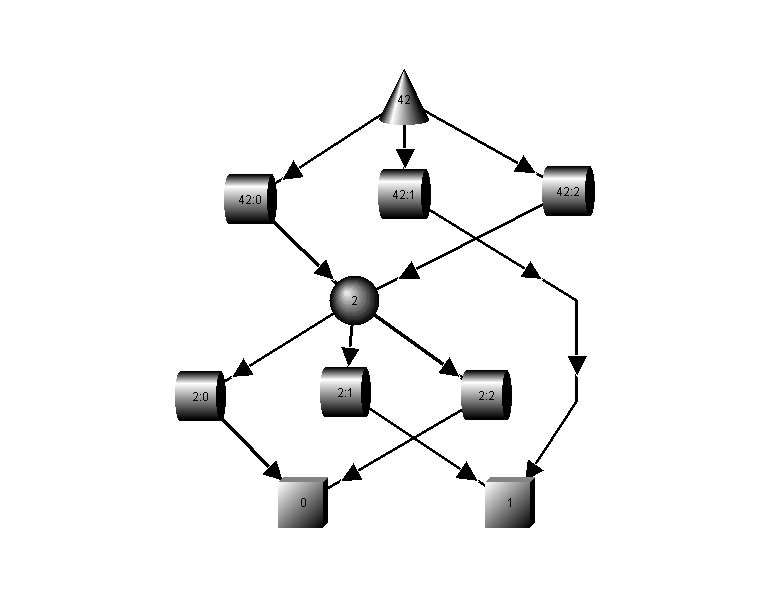}{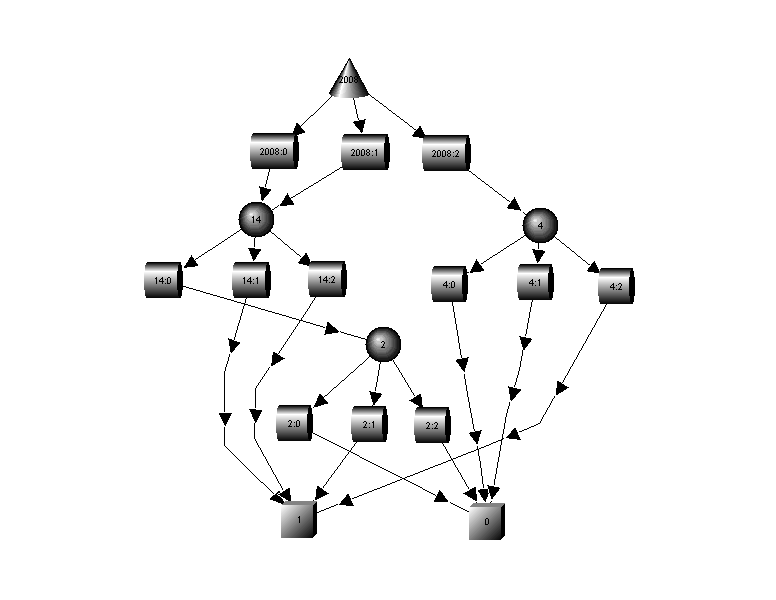}

The following proposition states that this tupling function is a 
generalization of {\tt bitmerge\_pair}

\begin{prop}
The following function equivalences hold:

\begin{equation}
bitmerge\_unpair~n \equiv to\_tuple~2~n
\end{equation}
\begin{equation}
bitmerge\_pair~(x,y) \equiv from\_tuple~[x,y]
\end{equation}
\end{prop}

\section{Encoding Finite Functions} \label{fun}

As finite sets can be put in a bijection with an initial segment 
of $Nat$ we can narrow down the concept of finite function as follows:
\begin{df}
A {\tt finite function} is a function defined from an initial 
segment of $Nat$ to $Nat$.
\end{df}

This definition implies that a finite function can be seen as an array or 
a list of natural numbers except that we do not limit the size of 
the representation of its values.

\subsection{Encoding Finite Functions as Tuples}

We can now encode and decode a finite function from $[0..k-1]$ to $Nat$ 
(seen as the list of its values), as a natural number:
\begin{code}  
ftuple2nat [] = 0
ftuple2nat ns = haskell2pepis (pred k,t) where
  k=genericLength ns 
  t=from_tuple ns

nat2ftuple 0 = []
nat2ftuple kf = to_tuple (succ k) f where 
  (k,f)=pepis2haskell kf
\end{code}
As the length of the tuple, {\tt k}, is usually smaller than the number 
obtained by merging the bits of the {\tt k}-tuple, we have picked the 
Pepis pairing function, exponential in its first argument and linear 
in its second, to embed the length of the tuple needed for the decoding. 
The encoding/decoding works as follows:
\begin{verbatim}
ftuple2nat [1,0,2,1,3]
  21295
nat2ftuple 21295
  [1,0,2,1,3]
map nat2ftuple [0..15]
  [[],[0,0],[1],[0,0,0],[2],[1,0],[3],
  [0,0,0,0],[4],[0,1],[5],[1,0,0],[6],
  [1,1],[7],[0,0,0,0,0]]
\end{verbatim}
Note that
\begin{verbatim}
map nat2ftuple [0..]
\end{verbatim}
\noindent provides an iterative generator for the stream of finite functions.

\subsection{Deriving an Encoding of Finite Functions from Ackermann's Encoding}

Given that a finite set with n elements can be put in a bijection 
with [0..n-1], a finite functions $f : [0..n-1] \rightarrow Nat$ 
can be represented as the list $[f(0)...f(n-1)]$. Such a list has 
however repeated elements. So how can we turn it into a set with 
distinct elements, bijectively?

The following two functions provide the answer.

First, we just sum up the list of the values of the function with {\tt scanl}, 
resulting in a monotonically growing sequence (provided that we first 
increment every number by 1 to ensure that 0 values do not break monotonicity).
\begin{code}
fun2set ns = 
  map pred (tail (scanl (+) 0 (map succ ns)))
\end{code}
The inverse function reverting back from a set of distinct values 
collects the increments from a term to the next (and ignores the last one):
\begin{code}
set2fun ns = map pred (genericTake l ys) where 
  l=genericLength ns
  xs =(map succ ns)
  ys=(zipWith (-) (xs++[0]) (0:xs))
\end{code}

\begin{prop}
The following function equivalences hold:

\begin{equation}
nat2set \circ set2nat \equiv id
\end{equation}

\begin{equation}
set2nat \circ nat2set \equiv id
\end{equation}
\end{prop}

The following example shows the conversion and its inverse.
\begin{verbatim}
fun2set [1,0,2,1,2]
  [1,2,5,7,10]
set2fun [1,2,5,7,10]
  [1,0,2,1,2]
\end{verbatim}

By combining this with Ackermann encoding's basic step {\tt set2nat} 
and its inverse {\tt nat2set}, we obtain an encoding 
from finite functions to $Nat$ follows:  
\begin{code}  
nat2fun = set2fun . nat2set

fun2nat = set2nat . fun2set
\end{code}
\begin{verbatim}
nat2fun 2008
  [3,0,1,0,0,0,0]
fun2nat [3,0,1,0,0,0,0]
  2008
\end{verbatim}

\begin{prop}
The following function equivalences hold:

\begin{equation}
nat2fun \circ fun2nat \equiv id
\end{equation}

\begin{equation}
fun2nat \circ nat2fun \equiv id
\end{equation}
\end{prop}

One can see that this encoding ignores {\tt 0}s in the binary representation 
of a number, while counting {\tt 1} sequences as increments. 
{\em Run Length Encoding} of binary sequences \cite{conf/cpm/MakinenN05} 
encodes {\tt 0}s and {\tt 1}s symmetrically, by counting the numbers 
of {\tt 1}s and {\tt 0}s. This encoding is reversible, 
knowing that {\tt 1}s and {\tt 0}s alternate, 
and that the most significant digit is always {\tt 1}:
\begin{code}
bits2rle [] = []
bits2rle [_] = [0]
bits2rle (x:y:xs) | x==y = (c+1):cs where 
  (c:cs)=bits2rle (y:xs)
bits2rle (_:xs) = 0:(bits2rle xs)

rle2bits [] = []
rle2bits (n:ns) = 
  (genericReplicate (n+1) b) ++ xs where 
    xs=rle2bits ns
    b=if []==xs then 1 else 1-(head xs)
\end{code}
By composing them with converters to/from bitlists, 
we obtain the bijection $nat2rle:Nat \rightarrow [Nat]$ 
and its inverse $rle2nat:[Nat] \rightarrow Nat$
\begin{code}
nat2rle = bits2rle . to_rbits0

rle2nat = from_rbits . rle2bits

to_rbits0 0 = []
to_rbits0 n = to_rbits n
\end{code}

\begin{prop}
The following function equivalences hold:

\begin{equation}
nat2rle \circ rle2nat \equiv id
\end{equation}

\begin{equation}
rle2nat \circ nat2rle \equiv id
\end{equation}
\end{prop}

\section{Encodings for ``Hereditarily Finite Functions"} \label{hff}
One can now build a theory of ``Hereditarily Finite Functions" 
($HFF$) centered around using a transformer 
like {\tt nat2ftuple}, {\tt nat2fun}, {\tt nat2rle} 
and its inverse {\tt ftuple2nat}, {\tt fun2nat}, {\tt rle2nat} 
in way similar to the use of {\tt nat2set} and {\tt set2nat} for $HFS$, 
where the empty function (denoted {\tt F []}) replaces the 
empty set as the quintessential {\em ``urfunction"}. 
Similarly to Urelements in the $HFS$ theory, ``urfunctions" 
(considered here as atomic values) can be introduced as constant functions 
parameterized to belong to $[0..ulimit-1]$.

By using the generic {\tt rank} function defined 
in section \ref{unrank} we can extend the bijections defined  in this 
section to encodings of Hereditarily Finite Functions.
By instantiating the transformer function in {\tt unrank} to {\tt nat2ftuple}, 
{\tt nat2fun} and {\tt nat2rle} we obtain:
\begin{code}

nat2hff = unrank nat2fun
nat2hff1 = unrank nat2ftuple
nat2hff2 =  unrank nat2rle
\end{code}

By instantiating the transformer function in {\tt rank} we obtain:
\begin{code}
hff2nat = rank fun2nat
hff2nat1 = rank ftuple2nat
hff2nat2 = rank rle2nat
\end{code}

The following examples show that {\tt nat2hff}, {\tt nat2hff1} 
and {\tt nat2hff2} are indeed bijections, and that the resulting 
$HFF$-trees are typically more compact than the $HFS$-tree 
associated to the same natural number. 
\begin{verbatim}
  F []
nat2hff 1
  F [F []]
nat2hff1 0
  F []
nat2hff1 1
  F [F [],F []]
nat2hff2 0
  F []
nat2hff2 1
  F [F []]

nat2hff 42
  F [F [F []],F [F []],F [F []]]
nat2hff1 42
  F [F [F [F [],F [],F []],F []]]
nat2hff2 42
  F [F [],F [],F [],F [],F [],F []]  
nat2hfs 42
  F [F [F []],F [F [],F [F []]],
     F [F [],F [F [F []]]]]    
  F [F [F []],F [F [],F [F []]],
     F [F [],F [F [F []]]]]

nat2hff 12345
  F [F [],F [F [F []]],F [],
     F [],F [F [F []],F []],F []] 
nat2hff1 12345
  F [F [F [F [F [F [],F []]],
    F []]],F [F [],F [],F [F [],F []]]]
nat2hff2 12345
  F [F [],F [F []],F [F [],F []],
    F [F [],F [],F []],F [F []]] 

hff2nat (nat2hff 12345)
  12345  
hff2nat1 (nat2hff1 12345)
  12345     
hff2nat2 (nat2hff2 12345)
  12345    
\end{verbatim}

Note that {\tt map nat2hff [0..], nat2hff1 [0..], nat2hff1 [0..]} 
provide iterative generators for the (recursively enumerable!) 
stream of hereditarily finite functions. 

The resulting HFF with urfunctions (seen as digits) can also 
be used as generalized {\em numeral systems} with applications 
to building arbitrary length integer implementations.
Assuming {\tt default\_ulimit=10} we obtain:
\begin{verbatim}
nat2hff 1234567890
  F [A 3,A 2,A 0,A 1,A 7,
     A 0,A 1,A 2,A 0,A 2,A 2
  ]
nat2hff1 1234567890
  F [F [F [F [F [A 0,A 3]],
     F [F [F [A 2,A 0,A 1]]],A 1]]
  ]
nat2hff2 1234567890
  F [A 2,A 0,A 1,A 1,A 0,A 0,A 6,A 1,
     A 0,A 0,A 1,A 1,A 1,A 0,A 1,A 0
  ]
\end{verbatim}
\noindent which display with the {\tt funShow} functions given in Appendix as:
\begin{verbatim}
funShow 1234567890
  "(3 2 0 1 7 0 1 2 0 2 2)"
funShow1 1234567890
  "(((((0 3)) (((2 0 1))) 1)))"
funShow2 1234567890
  "(2 0 1 1 0 0 6 1 0 0 1 1 1 0 1 0)"
\end{verbatim}
\begin{prop}
The following function equivalences hold:

\begin{equation}
nat2hff1 \circ hff2nat1 \equiv id
\end{equation}

\begin{equation}
hff2nat1 \circ nat2hff1 \equiv id
\end{equation}

\begin{equation}
nat2hff \circ hff2nat \equiv id
\end{equation}

\begin{equation}
hff2nat \circ nat2hff \equiv id
\end{equation}
\end{prop}

\section{Encoding Finite Bijections} \label{perm}
To obtain an encoding for finite bijections (permutations) 
we will first review a ranking/unranking mechanism for permutations that
involves an unconventional numeric representation, {\em factoradics}.

\subsection{The Factoradic Numeral System}
The factoradic numeral system \cite{knuth_art_1997-1} replaces digits
multiplied by power of a base $N$ with digits that multiply successive values
of the factorial of $N$. In the increasing order variant {\tt fr} the first
digit $d_0$ is 0, the second is $d_1 \in \{0,1\}$ and the $N$-th is $d_N \in
[0..N-1]$. The left-to-right, decreasing order variant {\tt fl} 
is obtained by reversing the digits of {\tt fr}.
\begin{verbatim}
fr 42
  [0,0,0,3,1]
rf [0,0,0,3,1]
  42
fl 42
  [1,3,0,0,0]
lf [1,3,0,0,0]
  42
\end{verbatim}
\noindent The function {\tt fr} handles the special case for $0$ and
calls {\tt fr1} which recurses and divides with increasing values of N
while collecting digits with {\tt mod}:
\begin{code}
-- factoradics of n, right to left
fr 0 = [0]
fr n = f 1 n where
   f _ 0 = []
   f j k = (k `mod` j) : 
           (f (j+1) (k `div` j))
\end{code}
The function {\tt fl}, with digits left to right is obtained as follows:
\begin{code}
fl = reverse . fr
\end{code}
The function {\tt lf} (inverse of {\tt fl}) converts back to decimals by
summing up results while computing the factorial progressively:
\begin{code}
rf ns = sum (zipWith (*) ns factorials) where
  factorials=scanl (*) 1 [1..]
\end{code}
Finally, {\tt lf}, the inverse of {\tt fl} is obtained as:
\begin{code}
lf = rf . reverse
\end{code}

\subsection{Ranking and unranking permutations of given size with Lehmer codes
and factoradics} 
The Lehmer code of a permutation $f$ is defined
as the number of indices j such that $1 \leq j < i$ and $f(j)<f(i)$
 \cite{DBLP:journals/dmtcs/MantaciR01}.
 \begin{prop}
 The Lehmer code of a permutation determines the permutation uniquely.
 \end{prop} 
The function {\tt perm2nth} computes a {\tt rank} 
for a permutation {\tt ps} of {\tt size>0}. 
It starts by first computing its Lehmer code {\tt ls} with 
{\tt perm2lehmer}. Then  it associates a unique natural 
number {\tt n} to {\tt ls}, 
by converting it with the function {\tt lf} 
from factoradics to decimals. 
Note that the Lehmer code {\tt Ls} is used as the list of digits
in the factoradic representation.
\begin{code}
perm2nth ps = (l,lf ls) where 
  ls=perm2lehmer ps
  l=genericLength ls

perm2lehmer [] = []
perm2lehmer (i:is) = l:(perm2lehmer is) where
  l=genericLength [j|j<-is,j<i]  
\end{code}

The function {\tt nat2perm} provides the matching {\em unranking}
operation associating a permutation {\tt ps} to a given {\tt size>0} 
and a natural number {\tt n}.
\begin{code}
-- generates n-th permutation of given size
nth2perm (size,n) = 
  apply_lehmer2perm (zs++xs) [0..size-1] where 
    xs=fl n
    l=genericLength xs
    k=size-l
    zs=genericReplicate k 0

-- converts Lehmer code to permutation   
lehmer2perm xs = apply_lehmer2perm xs is where 
  is=[0..(genericLength xs)-1]

-- extracts permutation from factoradic "digit" list    
apply_lehmer2perm [] [] = []
apply_lehmer2perm (n:ns) ps@(x:xs) = 
   y : (apply_lehmer2perm ns ys) where
   (y,ys) = pick n ps

pick i xs = (x,ys++zs) where 
  (ys,(x:zs)) = genericSplitAt i xs
\end{code}
Note also that {\tt lehmer2perm} is used this time
to reconstruct the permutation {\tt ps} from its Lehmer code,
which in turn is computed from the permutation's 
factoradic representation.

One can try out this bijective mapping as follows:
\begin{verbatim}
nth2perm (5,42)
  [1,4,0,2,3]
perm2nth [1,4,0,2,3]
  (5,42)
nth2perm (8,2008)
  [0,3,6,5,4,7,1,2]
perm2nth [0,3,6,5,4,7,1,2]
  (8,2008)
\end{verbatim}

\subsection{A bijective mapping from permutations to $Nat$}
One more step is needed to to extend the mapping between permutations of a
given length to a bijective mapping from/to $Nat$: we will have to ``shift
towards infinity'' the starting point of each new bloc of permutations in $Nat$
as permutations of larger and larger sizes are enumerated.

First, we need to know by how much - so we compute the sum of
all factorials up to $n!$.
\begin{code}
-- fast computation of the sum of all factorials up to n!
sf n = rf (genericReplicate n 1)
\end{code}
This is done by noticing that the factoradic representation of
[0,1,1,..] does just that. The stream of all such sums can now
be generated as usual:
\begin{code}
sfs = map sf [0..]
\end{code}
What we are really interested into, is decomposing {\tt n} into
the distance to the
last sum of factorials smaller than {\tt n}, {\tt n\_m}
and the its index in the sum, {\tt k}.
\begin{code}
to_sf n = (k,n-m) where 
  k=pred (head [x|x<-[0..],sf x>n])
  m=sf k
\end{code}
{\em Unranking} of an arbitrary permutation is now easy - the index {\tt k}
determines the size of the permutation and {\tt n-m} determines
the rank. Together they select the right permutation with {\tt nth2perm}.
\begin{code}
nat2perm 0 = []
nat2perm n = nth2perm (to_sf n)
\end{code}
{\em Ranking} of a permutation is even easier: we first compute
its size and its rank, then we shift the rank by 
the sum of all factorials up to its size, enumerating the
ranks previously assigned.
\begin{code}
perm2nat ps = (sf l)+k where 
  (l,k) = perm2nth ps
\end{code}
\begin{verbatim}
nat2perm 2008
  [1,4,3,2,0,5,6]
perm2nat [1,4,3,2,0,5,6]
  2008
\end{verbatim}
As finite bijections are faithfully represented by permutations,
this construction provides a bijection from $Nat$ to 
the set of Finite Bijections.
\begin{prop}
The following function equivalences hold:
\begin{equation}
nat2perm \circ perm2nat \equiv id \equiv perm2nat \circ nat2perm
\end{equation}
\end{prop}
The stream of all finite permutations can now
be generated as usual:
\begin{code}
perms = map nat2perm [0..]
\end{code}

\subsection{Hereditarily Finite Permutations}

By using the generic {\tt unrank} and {\tt rank} functions defined 
in section \ref{unrank} we can extend the {\tt nat2perm} and {\tt perm2nat} to
encodings of Hereditarily Finite Permutations ($HFP$).
\begin{code}
nat2hfp = unrank nat2perm
hfp2nat = rank perm2nat
\end{code}
The encoding works  as follows:
\begin{verbatim}
nat2hfp 42
  F [F [],F [F [],F [F []]],
    F [F [F []],F []],F [F []],
    F [F [],F [F []],F [F [],F [F []]]]]
hfp2nat it
  42
\end{verbatim}
\noindent Assuming {\tt default\_ulimit=10} and using the string representation
provided by permShow (Appendix) we obtain:
\begin{verbatim}
nat2hfp 42
  F [F [],A 2,A 3,A 1,A 4]
permShow 42
  "(0 2 3 1 4)"
permShow 1234567890
  "(1 6 (0 1 3 2) 2 0 3 (0 1 2 3) 
    7 8 5 9 4 (0 2 1 3))"
\end{verbatim}
\begin{prop}
The following function equivalences hold:
\begin{equation}
nat2hfp \circ hfp2nat \equiv id \equiv hfp2nat \circ nat2hfp
\end{equation}
\end{prop}

\section{Related work} \label{related}
Natural Number encodings of Hereditarily Finite Sets have 
triggered the interest of researchers in fields ranging from 
Axiomatic Set Theory and Foundations of Logic to 
Complexity Theory and Combinatorics
\cite{finitemath,kaye07,DBLP:journals/mlq/Kirby07,abian78,DBLP:journals/jsyml/Booth90,DBLP:journals/jct/MeirMM83,DBLP:conf/foiks/LeontjevS00,DBLP:journals/tcs/Sazonov93,avigad97}. 
Computational and Data Representation aspects of Finite Set Theory 
have been described in logic programming and theorem proving contexts 
in \cite{dovier00comparing,DBLP:journals/tplp/PiazzaP04,DBLP:conf/types/Paulson94}. 
Pairing functions have been used work on decision problems as early 
as \cite{pepis,kalmar1,robinson50,robinsons68b}. The tuple functions
we have used to encode finite functions are new. 
While finite functions have been used extensively in various branches of mathematics 
and computer science, we have not seen any formalization of hereditarily 
Finite Functions or Hereditarily Finite Bijections as such in the literature.

\section{Conclusion and Future Work} \label{concl}

We have shown the expressiveness of Haskell as a
metalanguage for executable mathematics, by describing
natural number encodings, tupling/untupling and ranking/unranking functions
for finite sets, functions and permutations and by extending them in a
generic way to Hereditarily Finite Sets, Hereditarily Finite Functions
and Hereditarily Finite Permutations.

In a Genetic Programming context \cite{koza92,poli08}, 
the bijections between bitvectors/natural numbers 
on one side, and trees/graphs representing HFSs, HFFs, HPPs on the other side, 
suggest exploring the mapping and its action on various transformations 
as a phenotype-genotype connection.

We also foresee interesting applications in cryptography and steganography. 
For instance, in the case of the permutation related encodings -  something as
simple as the order of the cities visited or the order of names 
on a greetings card, seen as a permutation with respect to their 
alphabetic order, can provide a steganographic encoding/decoding of a secret
message by using functions like {\tt nat2perm} and {\tt perm2nat}. It
looks like an interesting topic to investigate if higher density and more random
looking steganographic loads could be incorporated on top of Hereditarily Finite
Permutations. 

\bibliographystyle{plainnat}
\bibliography{INCLUDES/theory,tarau,INCLUDES/proglang,INCLUDES/biblio,INCLUDES/syn}

\appendix
\section{Appendix}
To make the code in the paper fully self contained, 
we list here some auxiliary functions.

\paragraph{String Representations}
The functions {\tt setShow} and {\tt funShow} provide a
string representation of a natural number as a ``pure" HFS or HFF.
They are obtained as instances of {\tt gshow} which provides a 
generic template parameterized with syntactic elements.
\begin{code}
setShow = (gshow "{" "," "}") . nat2hfs
funShow = (gshow "(" " " ")") . nat2hff
funShow1 = (gshow "(" " " ")") . nat2hff1
funShow2 = (gshow "(" " " ")") . nat2hff2
permShow = (gshow "(" " " ")") . nat2hfp

gshow _ _ _ (A n) = show n
gshow l _ r (F []) = 
  -- empty function shown as 0 rather than ()
  if default_ulimit > 1 then "0" else l++r
gshow l c r (F ns) = l++ 
  foldl (++) "" 
    (intersperse c (map (gshow l c r) ns)) 
  ++r  
\end{code}

\paragraph{Bit crunching functions} 
The function
bitcount computes the number of bits needed to represent an integer and
max\_bitcount computes the maximum bitcount for a list of integers.
\begin{code}
bitcount n = head [x|x<-[1..],(exp2 x)>n]
max_bitcount ns = foldl max 0 (map bitcount ns)
\end{code}

The following functions implement conversion operations
 between bitlists and numbers.
Note that our bitlists represent binary numbers by 
selecting exponents of 2 in 
increasing order (i.e. ``right to left"). 
\begin{code}
-- from decimals to binary as list of bits
to_rbits n = to_base 2 n

-- from bits to decimals
from_rbits bs = from_base 2 bs

-- to binary, padded with 0s, up to maxbits
to_maxbits maxbits n = 
  bs ++ (genericTake (maxbits-l)) (repeat 0) where 
    bs=to_base 2 n
    l=genericLength bs

-- conversion to base n, as list of digits
to_base base n = d : 
  (if q==0 then [] else (to_base base q)) where
    (q,d) = quotRem n base

-- conversion from any base to decimal 
from_base base [] = 0
from_base base (x:xs) = x+base*(from_base base xs)
\end{code}

\end{document}

%% file: TOOLS/hheader.tex
\usepackage{amsmath}
\usepackage{graphicx}
\usepackage{subfig}
\usepackage{url}
\usepackage{verbatim} 
\usepackage{listings}
\lstnewenvironment{codex}
    {\lstset{}%
      \csname lst@SetFirstLabel\endcsname}
    {\csname lst@SaveFirstLabel\endcsname}
\lstnewenvironment{code}
    {\lstset{}%
      \csname lst@SetFirstLabel\endcsname}
    {\csname lst@SaveFirstLabel\endcsname}
\newtheorem{prop}{Proposition}
\newtheorem{df}{Definition}

\newcommand{\BI}[0]{\begin{itemize}}
\newcommand{\EI}[0]{\end{itemize}}

\newcommand{\BE}[0]{\begin{enumerate}}
\newcommand{\EE}[0]{\end{enumerate}}

\newcommand{\BX}[0]{\begin{codex}}
\newcommand{\EX}[0]{\end{codex}}
\newcommand{\BV}[0]{\begin{verbatim}}
\newcommand{\EV}[0]{\end{verbatim}}
    
\def \bscale1 {0.50}
\def \bscale {0.25}

\newcommand{\FIG}[3]{
\begin{figure}[htbp]
\centering
\scalebox{\bscale1}{\includegraphics*[bb=0pt 0pt 800pt 800pt]{./figs/#3.png}}
\caption{#2}
\label{#1}
\end{figure}
}

\newcommand{\VFIGS}[6]{
\begin{figure}[htbp]
  \begin{center}
    {\scalebox{\bscale}{\includegraphics*[bb=0pt 0pt 800pt 800pt]{./figs/#5.png}}}
    {\scalebox{\bscale}{\includegraphics*[bb=0pt 0pt 800pt 800pt]{./figs/#6.png}}}
  \caption{#2: {\em #3} and {\em #4}}
  \label{#1}
  \end{center}
\end{figure}
}